\documentclass[10pt,twocolumn,letterpaper]{article}

\usepackage{amsmath,amsfonts,bm}

\def\eqref#1{equation~\ref{#1}}

\def\1{\bm{1}}

\def\vz{{\bm{z}}}

\DeclareMathAlphabet{\mathsfit}{\encodingdefault}{\sfdefault}{m}{sl}
\SetMathAlphabet{\mathsfit}{bold}{\encodingdefault}{\sfdefault}{bx}{n}

\def\sB{{\mathbb{B}}}

\usepackage{iccv}
\usepackage{times}
\usepackage{epsfig}
\usepackage{graphicx}
\usepackage{amsmath}
\usepackage{amssymb}
\usepackage{multirow}
\usepackage{float}
\usepackage{mathbbol}

\usepackage[utf8]{inputenc} 
\usepackage[T1]{fontenc}    
\usepackage{url}            
\usepackage{booktabs}       
\usepackage{amsfonts}       
\usepackage{nicefrac}       
\usepackage{microtype}      
\usepackage{xcolor}         
\usepackage{graphicx}
\usepackage{wrapfig}

\usepackage{makecell}

\usepackage{algorithm}  
\usepackage{algorithmicx}  
\usepackage{algpseudocode} 
\usepackage{paralist}
\usepackage{bbm}

\usepackage[pagebackref=true,breaklinks=true,letterpaper=true,colorlinks,bookmarks=false]{hyperref}

\usepackage[capitalize]{cleveref}
\crefname{section}{Sec.}{Secs.}
\Crefname{section}{Section}{Sections}
\Crefname{table}{Table}{Tables}
\crefname{table}{Tab.}{Tabs.}

\iccvfinalcopy 

\def\iccvPaperID{9336} 

\ificcvfinal\fi

\begin{document}

\title{Robust and IP-Protecting Vertical Federated Learning against\\
Unexpected Quitting of Parties}

\author{Jingwei Sun$^1$, Zhixu Du$^1$, Anna Dai$^1$, Saleh Baghersalimi$^2$, Alireza Amirshahi$^2$, \\David Atienza$^2$, Yiran Chen$^1$\\
$^1$ Department of Electrical and Computer Engineering, Duke University\\
$^2$ École Polytechnique Fédérale de Lausanne (EPFL)\\
{\tt\small $^1$\{jingwei.sun, zhixu.du, anna.dai, yiran.chen\}@duke.edu,}\\
{\tt\small $^2$\{saleh.baghersalimi, alireza.amirshahi, david.atienza\}@epfl.ch}
}

\maketitle
\ificcvfinal\thispagestyle{empty}\fi

\begin{abstract}
    Vertical federated learning (VFL) enables a service provider (i.e., active party) who owns labeled features to collaborate with passive parties who possess auxiliary features to improve model performance. Existing VFL approaches, however, have two major vulnerabilities when passive parties unexpectedly quit in the deployment phase of VFL - severe performance degradation and intellectual property (IP) leakage of the active party's labels. In this paper, we propose \textbf{Party-wise Dropout} to improve the VFL model's robustness against the unexpected exit of passive parties and a defense method called \textbf{DIMIP} to protect the active party's IP in the deployment phase. We evaluate our proposed methods on multiple datasets against different inference attacks. The results show that Party-wise Dropout effectively maintains model performance after the passive party quits, and DIMIP successfully disguises label information from the passive party's feature extractor, thereby mitigating IP leakage.


\end{abstract}

\section{Introduction}

Federated learning (FL)~\cite{mcmahan2017communication,yang2019federated} is a distributed learning method that allows multiple parties to collaboratively train a model without directly sharing their data, thereby preserving their data privacy. FL was initially proposed as Horizontal Federated Learning (HFL) to enable collaborative learning across devices~\cite{sun2022fedsea}. In this case, data is "horizontally" split, where the devices share the same feature space but have different samples. Another FL framework is Vertical Federated Learning (VFL), which focuses on scenarios where various parties have data with different feature spaces but share overlapping samples~\cite{liu2019communication,liu2021fate}. Different from HFL, VFL is mostly deployed in cross-silo scenarios. Suppose a service provider, referred to as the active party, owns data and labels of its clients and wishes to train a deep learning model. The active party may collaborate with other parties (i.e., passive parties) that possess different data features of the same clients to boost the model's performance. Instead of explicitly sharing the raw data, the passive parties transmit the extracted representations from their feature extractors to the active party for training and inference.

\begin{figure}[ht]
\centering
     \includegraphics[scale=0.5]{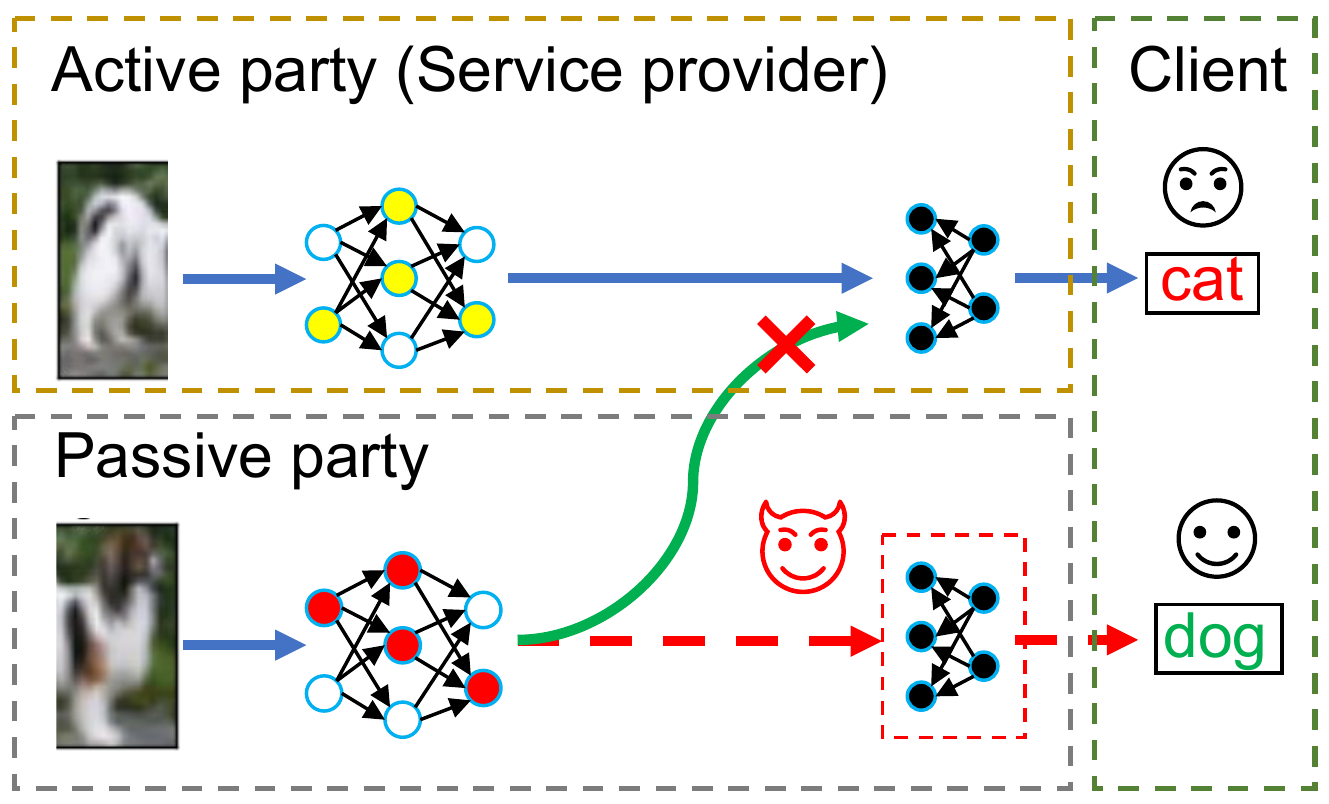}
\caption{The passive party might quit in the deployment phase, which would cause a substantial performance drop. The passive party could also extract representations containing the information of the active party's labels using its feature extractor, leading to IP leakage of the active party.}
\label{fig:intro}
\end{figure}

The collaboration of devices in HFL happens in the training phase, and the global model is deployed on each device for local inference in the deployment phase. In contrast, VFL requires the parties to collaborate in both the training and deployment phases. During the deployment phase, the active party still requires the representations uploaded by passive parties to conduct inference. However, in real-world scenarios, it is possible for passive parties to quit unexpectedly at inference time due to network crashes, system maintenance, or termination of collaborations. When unexpected quitting happens, the service provider faces two challenges: (1) a substantial \textbf{performance drop}; (2) potential \textbf{intellectual property (IP) leakage} through the passive party's feature extractor. This paper shows that the drop in model performance caused by the passive party's quitting results in a model that performs worse than one trained by the active party alone, ultimately undermining the motivation for VFL. Furthermore, the passive parties can retain access to their feature extractors even after terminating the collaboration. These feature extractors are trained using the active party's labels, which are valuable IP. From these feature extractors, the passive parties can extract representations containing the information of the active party's labels. Although previous studies made efforts towards mitigating label information leakage through inference attacks on gradients during the training phase~\cite{chaudhuri2011sample, ghazi2021deep}, the robustness and IP protection of VFL in the deployment phase remain under-explored. 

In this paper, we show that the substantial performance drop after a passive party quits is caused by the co-adaptation of feature extractors across parties. Co-adaptation refers to the phenomenon where the classifier relies heavily on certain representation components from specific parties' extractors. We propose \textbf{Party-wise Dropout} as a solution to this co-adaptation problem to alleviate the performance drop when passive parties quit unexpectedly. To prevent the IP leakage of the active party's labels, we propose a defense called \textbf{DIMIP}\footnote{\textbf{D}efense against \textbf{I}P Leakage from \textbf{M}utual \textbf{I}nformation \textbf{P}erspective} that minimizes the mutual information (MI) between the representations of the passive party and the true labels. We formulate DIMIP into an adversarial training algorithm that jointly minimizes the variational MI upper bound and prediction loss.

Our key contributions are summarized as follows:
\begin{compactitem}
    \item We investigate the reason behind the substantial drop in model performance after the unexpected quitting of passive parties in VFL and propose Party-wise Dropout to mitigate this performance drop.
    \item We propose a defense method, DIMIP, against the IP leakage of the active party's labels from the mutual information perspective with negligible degradation of the model's utility. 
    \item We empirically evaluate the performance of Party-wise Dropout and DIMIP with different datasets. Our results show that Party-wise Dropout can improve the accuracy after the passive party's exit by more than 8\% on CIFAR10. DIMIP prevents the passive party from fine-tuning a classifier that outperforms random guess levels even using the entire labeled dataset with only less than 2\% drop in the VFL model accuracy, outperforming baselines significantly.
\end{compactitem}

\section{Related Work}
\subsection{Vertical Federated Learning}
Vertical federated learning (VFL) has been an emerging research area since proposed. In contrast to (horizontal) federated learning (HFL), VFL adopts a different scheme for data partitioning~\cite{hardy2017private, yang2019federated}. In VFL, different parties will have various parts of the data of an overlapping individual. 
There has been an amount of research devoted to VFL. Specifically, \cite{hardy2017private} proposes a protocol involving a trusted third party to manage the communication utilizing homomorphic encryption, with the following works~\cite{hardy2017private, nock2018entity} on protocols design. 
Others have been following \cite{hardy2017private}, where \cite{nock2018entity} is working on assessing the protocols and \cite{yang2019quasi, yang2019parallel} are focusing on algorithm design concerning optimization. 
Additionally, VFL algorithms on traditional machine learning, such as tree-boosting~\cite{cheng2021secureboost}, gradient boosting~\cite{wu2020privacy, fu2021vf2boost}, random forest~\cite{liu2020federated}, linear regression~\cite{zhang2021secure}, and logistic regression~\cite{hu2019learning, liu2019communication} are also proposed. Another line of research is working on communication efficiency~\cite {liu2019communication}, which decreases the communication frequency by leveraging stale gradients on local training. Besides, the assumption of overlapping individuals in VFL among parties produces a challenge for applying VFL in the real world, where FedMVT~\cite{kang2020fedmvt} proposes to estimate representations and labels to alleviate the gap. 
Other efforts have also been made to apply VFL in the real world. For example, FATE~\cite{liu2021fate} is an open-source platform for building the end-to-end system. 

\subsection{IP Leakage in VFL}
Intellectual Property (IP) is drawing more and more attention as the rapid growth of commercial deployment of deep learning, especially in federated learning scenarios, whose primary concern is privacy. IP leakage can be divided into data IP leakages, such as deep leakage from gradients (DLG)~\cite{zhu2019deep, zhao2020idlg}, model inversion~\cite{fredrikson2015model} and their variants~\cite{geiping2020inverting, jin2021cafe, yin2021see, melis2019exploiting, jiang2022comprehensive}, and model IP leakage, such as model extraction attacks~\cite{tramer2016stealing, orekondy2019knockoff, pal2019framework, correia2018copycat, truong2021data}, where multiple defensive methods have also been proposed to tackle data IP leakage~\cite{so2020byzantine, mo2021ppfl, abadi2016deep, bonawitz2017practical} and model IP leakage~\cite{juuti2019prada, orekondy2019prediction}.

In VFL, we categorize IP stealing attacks into two types, i.e., feature inference~\cite{luo2021feature, he2019model, jiang2022comprehensive, jin2021cafe} and label inference~\cite{fu2022label, li2021label, liu2021defending}. 
Specifically, \cite{luo2021feature} proposes general attack methods for complex models, such as Neural Networks, by matching the correlation between adversary features and target features, which can be seen as a variant of model inversion~\cite{fredrikson2015model,sun2021soteria}. \cite{he2019model, jiang2022comprehensive} also propose variants of model inversion attack in VFL. While all these attacks are in the inference phase, \cite{jin2021cafe} proposes a variant of DLG~\cite{zhu2019deep} which can perform attacks in the training phase. 
For label inference, \cite{li2021label} proposes an attack method and a defense method for two-party split learning on binary classification problems, a special VFL setting. Additionally, \cite{fu2022label} proposes three different label inference attack methods considering different settings in VFL: direct label inference attack, passive label inference attack, and active label inference attack. 
Defensive methods have also been proposed. For example, \cite{liu2021defending} proposes manipulating the labels following specific rules to defend the direct label inference attack, which can be seen as a variant of label differential privacy (label DP)~\cite{chaudhuri2011sample, ghazi2021deep} in VFL. However, all these defending methods focus on preventing data IP leakage from gradients in the training phase. 
To the best of our knowledge, we are the first to provide an analysis of label IP protection in the VFL deployment phase.

\section{Problem Definition and Motivation}

\subsection{Vertical Federated Learning Setting}\label{sec:problem_def}

Suppose $K$ parties train a model. There is a dataset\footnote{We assume the alignment between overlapping samples is known as a prior. In some applications, private set intersection could be used before running VFL to find the sample alignment.} across all parties with size $N$:  $D=\{x_{i}, y_{i}\}_{i=1}^{N}$. The feature vector $x_{i}\in \mathbb{R}^{d}$ is distributed among $K$ parties $\{x_{i}^{k}\in \mathbb{R}^{d_k}\}_{k=1}^K$, where $d_k$ is the feature dimension of party $k$, and the labels $Y=\{y_{i}\}_{i=1}^{N}$ are owned by one party. The parties with only features are referred to as \textit{passive parties}, and the party with both features and labels is referred to as the \textit{active party}. Without loss of generality, we assume party $1$ is the active party, and other parties are passive parties. 

Each party (say the $k$-th) adopts a representation extractor $f_{\theta_k}(.)$ to extract representations of local data $H^k=\{H_i^k\}_{i=1}^N=\{f_{\theta_k}(x_i^k)\}_{i=1}^N$ and sends them to the active party for loss calculation. The overall training objective of VFL is formulated as

{\small
\begin{equation}
    \min\limits_{\Theta} \mathcal{L}(\Theta;D) \triangleq \frac{1}{N} \sum_{i=1}^{N} \mathcal{L}\left(\mathcal{S}_{\theta_{\mathcal{S}}}\left(H_i^1, ..., H_i^K\right), y_i\right),
    \label{eq:problem_def}
\end{equation}
}%

\noindent where $\Theta = \left[ \theta_1;...;\theta_K;\theta_{\mathcal{S}} \right]$, $\mathcal{S}$ denotes a trainable head model adopted by the active party to conduct classification based on the received representations, and $\mathcal{L}$ denotes the loss function. The objective of each passive party $k$ is to find the optimal $\theta_k^*$ without sharing local data $\{x_{i}^k\}_{i=1}^{N}$ and parameters $\theta_k$. The objective of the active party is to optimize $\theta_1$ and $\theta_{\mathcal{S}}$ without sharing $\theta_1$, $\theta_{\mathcal{S}}$ and true labels $Y$. To perform training, the active party conducts back-propagation to calculate the gradients of received representations $\{\frac{\partial \mathcal{L}}{\partial H^k}\}_{k\in[K]}$ and send $\{\frac{\partial \mathcal{L}}{\partial H^k}\}_{k\in[2,...,K]}$ back to corresponding parties.

After the training phase, the active and passive parties collaborate to conduct inference of new data samples. For a new data $x_i$, the passive parties send the extracted representations $\{H_i^k\}_{k\in[2,..., K]}$ to the active party, and the active party generates the prediction $\mathcal{S}_{\theta_{\mathcal{S}}}\left(H_i^1, ..., H_i^K\right)$. Notably, the passive parties still have to communicate with the active party during the inference phase and could fail transmission.

\subsection{Performance drop after parties quit}\label{sec:performance_drop}

During the deployment phase, some passive parties (say the $k$-th party) could quit unexpectedly due to a network crash or the termination of collaboration. Without the representations uploaded by party $k$, the active party can still conduct inference by setting $H^k_i$ as a zero vector. However, there will be a substantial performance drop. We conduct two-party experiments on CIFAR10 to investigate this performance drop. We follow previous works~\cite{liu2019communication,kang2022fedcvt} to split CIFAR10 images into two parts and assign them to the two parties using ResNet18 as backbone models. The active party (party 1) and passive party (party 2) collaborate to train the models. We evaluate and compare the inference accuracy before and after party 2 quits in the deployment phase. When party 2 quits, party 1 sets $H^2_i$ as a zero vector and conducts inference. Zero vectors are used because the passive party typically does not allow the active party to utilize its representations in any way (e.g., an average vector) after the termination of collaboration. We set the standalone results as a baseline, where the active party trains a model independently without ever collaborating with the passive party. The results are shown in \cref{tb:quit_prelim}.

\begin{table}[th]
\small
    \centering
    \caption{Compared results before and after party 2 quits on CIFAR10.}
        \begin{tabular}{l | c }
            \toprule
            & Accuracy(\%)\\
            \hline
            Before party 2 quits & 74.53\\
            \hline
            After party 2 quits & 51.24\\
            \hline
            Party 1 standalone & 62.84\\
            \bottomrule
        \end{tabular}
    \label{tb:quit_prelim}
\end{table}

The results show that the accuracy drops more than 20\% after party 2 quits. Furthermore, the VFL model after party 2 quits achieves even lower accuracy than the model party 1 trained without any collaboration, undermining the motivation of VFL.

\subsection{IP leakage of labels in the deployment phase}

The collaborative training process enables passive parties to extract representations useful for the task of VFL, which is defined by and learned from the labels of the active party. Even after the collaboration ends, the passive parties will retain access to the representation extractors since they conduct the training and deployment of the extractors. These extractors allow the passive parties to fine-tune classifier heads with very few labeled data and conduct inference with decent accuracy after quitting the collaboration. Given the active party's significant investment of effort and money in labeling the data, these extractors retained by the passive parties constitute costly IP leakage of these labels. 
To demonstrate the extent of IP leakage by the feature extractors of the passive parties, we follow the experimental setup in \cref{sec:performance_drop} and let party 2 conduct model completion (MC) attack~\cite{chong2022baseline} to train a classifier using a small number of labeled samples. We report the test accuracy of the complete model of party 2 created by the MC attack. For comparison, we also assume party 2 annotates all the training data to train a model from scratch and we report the accuracy in \cref{tb:ip_prelim}.

\begin{table}[th]
\small
    \centering
    \caption{Compared accuracy of the model on party 2 by conducting MC attack and collecting labels to train from scratch.}
        \begin{tabular}{l | c }
            \toprule
            & Accuracy(\%)\\
            \hline
            MC attack (400 labels) & 58.02\\
            \hline
            Train from scratch w. all the labels & 59.73\\
            \bottomrule
        \end{tabular}
    \label{tb:ip_prelim}
\end{table}

By fine-tuning a classifier with the extractor, the passive party can achieve comparable accuracy using less than 1\% of the labeled data compared to training a model from scratch with all the labels. This demonstrates that the label information from the active party is leaked and embedded in the passive party's extractor.

\section{Methods}

\subsection{Party-wise Dropout}

The severe performance drop after the quitting of passive parties comes from the co-adaptation of the hidden units of the head predictor $\mathcal{S}_{\theta_{\mathcal{S}}}$ and the neurons of local extractors $f_{\theta_k}$, which means that a hidden unit of the predictor $\mathcal{S}_{\theta_{\mathcal{S}}}$ is only dependent on the context of several specific neurons of specific parties' extractors. The co-adaptation problem was also studied in the deep neural networks~\cite{baldi2013understanding}, and \textit{dropout} was proposed as an effective solution. Enlightened by dropout, we propose \textbf{Party-wise Dropout}, which can solve the party-wise co-adaptation problem in VFL. For each round of communication (i.e., an iteration of training in VFL), the active party randomly omits the representations uploaded by the passive parties with probabilities of $[p^2,...,p^K]$, so the hidden units of the predictor do not rely only on any specific passive party's representations. 

\paragraph{Multi-objective training} Suppose an active party (party 1) and several passive parties ($\{\text{party } k\}_{k\in [2,...,K]}$) collaborate in VFL. The active party applies Party-wise Dropout to omit party $k$ with probability $p^k$. Then the expected training loss can be formulated as



\begin{equation}
    \begin{aligned}
        &\mathbb{E}_{p^2, ..., p^K}\left[ \mathcal{L}(\Theta;D)\right] \\ 
        = &\sum_{\vz=(z^2, ..., z^K) \in \{0,1\}^{K-1}} \prod_{k=2}^K (p^k)^{z^k}(1-p^k)^{(1-z^k)} \ell^{\vz},
    \end{aligned}
    \label{eq:dropout}
\end{equation}
where 
\begin{equation*}
    \ell^\vz = \sum_{i=1}^N \mathcal{L}(\mathcal{S}_{\theta_{\mathcal{S}}}(H_i^1, \mathbbm{1}_{\{z^2=0\}}H_i^2,..., \mathbbm{1}_{\{z^K=0\}}H_i^K), y_i).
\end{equation*}

The term $\mathcal{L}(\mathcal{S}_{\theta_{\mathcal{S}}}(H_i^1, H_i^2,..., H_i^K), y_i)$ is the loss function to make predictions based on the representations of all $K$ parties, which is also the loss without applying Party-wise Dropout. $\mathcal{L}(\mathcal{S}_{\theta_{\mathcal{S}}}(H_i^1, 0,..., 0), y_i)$ is the loss based only on party 1's representations, which is the prediction after all passive parties quit. Thus, by applying Party-wise Dropout, the active party implicitly conducts multi-objective training, which includes the loss of any set of passive parties and can preserve performance when any passive party quits. Notably, a larger $p^k$ sets a larger weight for $\mathcal{L}(\mathcal{S}_{\theta_{\mathcal{S}}}(H_i^1,..., H_i^{k-1}, 0, H_i^{k+1},..., H_i^K), y_i)$. Thus, $p^k$ can be chosen based on the probability that party $k$ would quit. 


\subsection{Defense against IP leakage from mutual information perspective (DIMIP)}

Without loss of generality, we formulate our defense named DIMIP in the two-party scenario. Suppose the passive party (party 2) and active party (party 1) have sample pairs $\left\{\left(x_i^1, x_i^2, y_i\right)\right\}_{i=1}^N$ drawn from a distribution $p\left(x^1, x^2, y\right)$, and the representations of party $k$ is calculated as $h^k=f_{\theta_k}(x^k)$. We use $h^k$, $x^k$ and $y$ here to represent random variables, while $H_i^k$, $x_i^k$ and $y_i$ stand for deterministic values. Then our learning algorithm is to achieve two goals:

\begin{itemize}
    \item Goal 1: To preserve the performance of VFL, the main objective loss should be minimized.
    \item Goal 2: To reduce the IP leakage of labels from party 2, $\theta_2$ should not be able to extract representations $h^2$ containing much information about the true label $y$.
\end{itemize}

\noindent Formally, we have two training objectives:

\begin{equation}
    \begin{aligned}
    &\textbf{Prediction:} \min\limits_{\theta_1, \theta_2, \theta_S} \mathcal{L}\left(\mathcal{S}_{\theta_{\mathcal{S}}}\left(h^1, h^2\right), y\right),\\
    &\textbf{Label IP protection:} \min\limits_{\theta_2} \text{I}(h^2;y),
    \end{aligned}
\end{equation}

\noindent where $\text{I}(h^2;y)$ is the mutual information between $h^2$ and $y$, which indicates the information $h^2$ preserves for the label variable $y$. We minimize this mutual information to protect the active party's labels' IP from being learned by the passive party. Ideally, if the $\text{I}(h^2;y)=0$, no label can be inferred by the passive party.

The prediction objective is usually easy to calculate (e.g., cross-entropy loss for classification). The mutual information term is hard to compute in practice as the random variable $h^2$ is high-dimensional. In addition, the mutual information requires knowing the distribution $p(y|h^2)$, which is difficult to compute. To derive a tractable estimation of the mutual information objective, we leverage \textit{CLUB}~\cite{cheng2020club} to formulate a variational upper-bound:

{\small
\begin{equation}
    \begin{aligned}
        &\text{I}\left(h^2;y\right)\\
        \leq~& \text{I}_\text{{vCLUB}}\left(h^2;y\right)\\
        :=~& \mathbb{E}_{p\left(h^2, y\right)}\log q_{\psi}\left(y|h^2\right)-\mathbb{E}_{p\left(h^2\right)p\left(y\right)}\log q_{\psi}\left(y|h^2\right),
    \end{aligned}
    \label{eq:vclub}
\end{equation}
}%

\noindent where $q_{\psi}\left(y|h^2\right)$ is a variational distribution with parameters $\psi$ to approximate $p\left(y|h^2\right)$. To reduce the computational overhead of the defense, we apply the sampled vCLUB (vCLUB-S) MI estimator in \cite{cheng2020club}, which is an unbiased estimator of $\text{I}_\text{{vCLUB}}$ and is formulated as
{\small
\begin{equation}
    \hat{\text{I}}_{\text{vCLUB-S}}(h^2;y) = \frac{1}{N} \sum\limits_{i=1}^N \left[\log q_{\psi}\left(y_i|H_i^2\right) - \log q_{\psi}\left(y_{k_i'}|H_i^2\right)\right],
\label{eq:vclubs}
\end{equation}
}%

\noindent where $k_i'$ is uniformly sampled from indices $\{1,...,N\}$. It is notable that to guarantee the first inequality of \cref{eq:vclub}, $q_{\psi}\left(y|h^2\right)$ should satisfy 

{\small
\begin{equation}
    \text{KL}\left(p\left(h^2,y\right)||q_\psi\left(h^2,y\right)\right)\leq \text{KL}\left(p\left(h^2\right)p\left(y\right)||q_\psi\left(h^2,y\right)\right),
\end{equation}
}
\noindent which can be achieved by minimizing $\text{KL}\left(p\left(h^2,y\right)||q_\psi\left(h^2,y\right)\right)$:

{\small
\begin{equation}
    \begin{aligned}
        &\min\limits_\psi \text{KL}\left(p\left(h^2,y\right)||q_\psi\left(h^2,y\right)\right)\\
        =~&\min\limits_\psi \mathbb{E}_ {p\left(h^2,y\right)}\left[\log\left(p\left(y|h^2\right)p\left(h^2\right)\right)-\log\left(q_{\psi}\left(y|h^2\right)p\left(h^2\right)\right)\right]\\
        =~&\min\limits_\psi \mathbb{E}_ {p\left(h^2,y\right)}\left[\log\left(p\left(y|h^2\right)\right)-\log\left(q_{\psi}\left(y|h^2\right)\right)\right].
    \end{aligned}
\end{equation}
}%

\noindent Since the first term has no relation to $\psi$, we just need to minimize $\mathbb{E}_ {p\left(h^2,y\right)}-\log\left(q_{\psi}\left(y|h^2\right)\right)$. With samples $\left\{\left(x_i^1, x_i^2, y_i\right)\right\}_{i=0}^N$, we can derive an unbiased estimation 

{\small
\begin{equation}
    \max\limits_{\psi}\frac{1}{N} \sum\limits_{i=1}^N \log q_{\psi}\left(y_i|H_i^2\right).
\label{eq:condition}
\end{equation}
}%

\noindent With \cref{eq:vclub}, \cref{eq:vclubs} and \cref{eq:condition}, the objective of label IP protection can be achieved by optimizing

{\small
\begin{equation}
    \begin{aligned}
        &\min\limits_{\theta_2} \text{I}(h^2;y)\\
        \Leftrightarrow~&\min\limits_{\theta_2} \hat{\text{I}}_{\text{vCLUB-S}}(h^2;y)\\
        =~&\min\limits_{\theta_2} \frac{1}{N} \sum\limits_{i=1}^N \left[\max\limits_{\psi}\log q_{\psi}\left(y_i|H_i^2\right) - \log q_{\psi}\left(y_{k_i'}|H_i^2\right)\right].
    \end{aligned}
    \label{eq:variational_objective}
\end{equation}
}%

\noindent Suppose we use $g_{\psi}$ to parameterize $q_{\psi}$, by combining \cref{eq:variational_objective} and the prediction objective with a weight hyper-parameter $\lambda$, we formulate the overall optimizing objective as


{\small
\begin{equation}
    \begin{aligned}
        &\min\limits_{\theta_1, \theta_2, \theta_S}(1-\lambda)\underbrace{\frac{1}{N} \sum_{i=1}^{N} \mathcal{L} \left(\mathcal{S}_{\theta_{\mathcal{S}}}\left(f_{\theta_1}\left(x_i^1\right),f_{\theta_2}\left(x_i^2\right)\right), y_i\right)}_{\mathcal{L}_C}\\
        &+\min\limits_{\theta_2}\max\limits_{\psi} \lambda\underbrace{\frac{1}{N} \sum\limits_{i=1}^N \log g_{\psi}\left(y_i|f_{\theta_2}\left(x_i^2\right)\right)}_{\mathcal{L}_A}\\
        &+\min\limits_{\theta_2} \lambda\underbrace{\frac{1}{N} \sum\limits_{i=1}^N-\log g_{\psi}\left(y_{k_i'}|f_{\theta_2}\left(x_i^2\right)\right)}_{\mathcal{L}_R}.
        \label{eq:final_object}
    \end{aligned}
\end{equation}
}%


\noindent It is notable that our defense is not limited to any specific type of task. For the classification problem, $-\mathcal{L}_A$ and $\mathcal{L}_R$ are equal to cross-entropy losses. 

\begin{figure}[th]
\centering
     \includegraphics[scale=0.32]{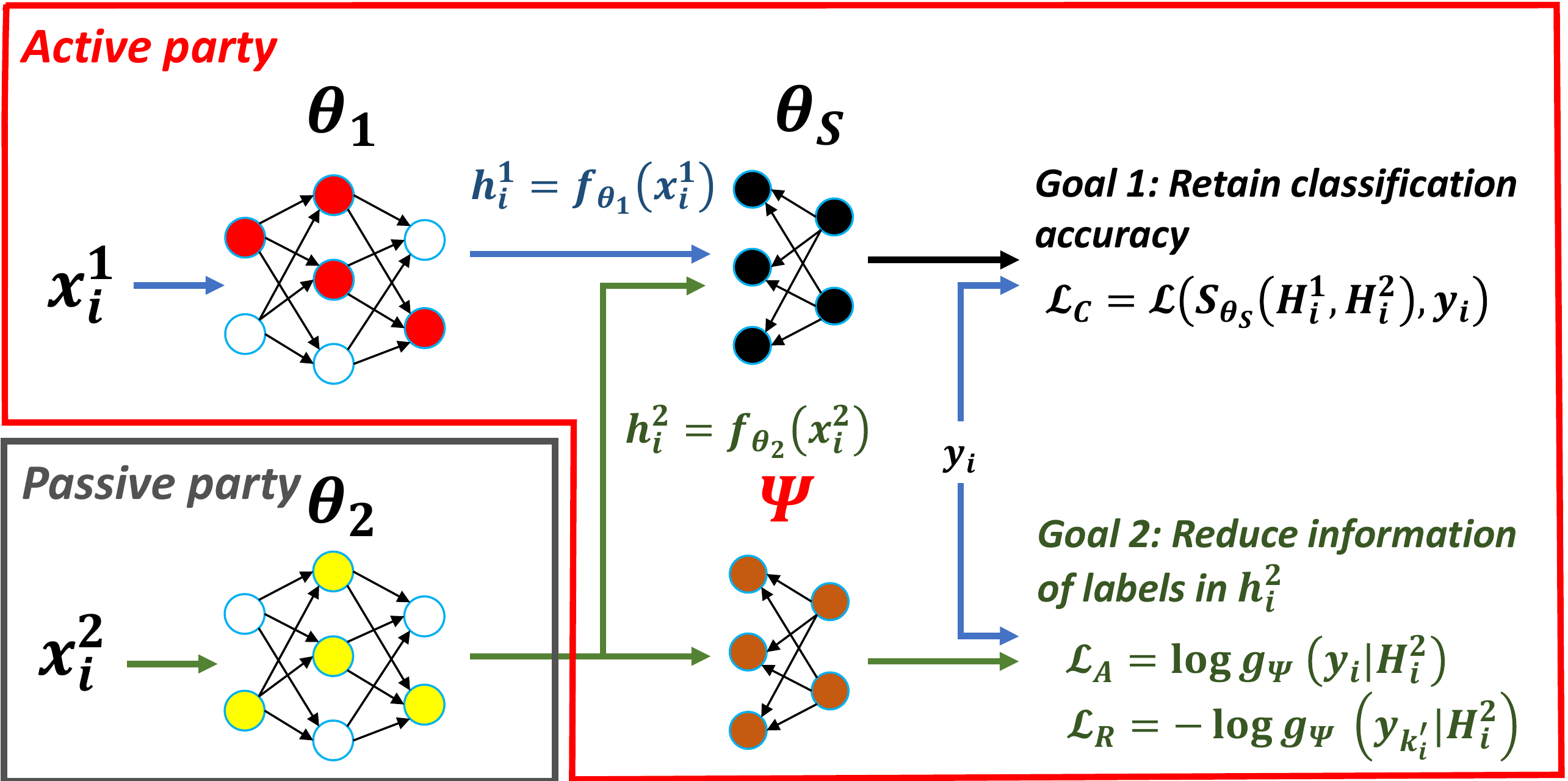}
\caption{Overview of DIMIP: The active party protects the IP of labels by optimizing $\mathcal{L}_A$ and $\mathcal{L}_R$ to reduce label information in passive parties' representations.}
\label{fig:DIMIP}
\end{figure}

\paragraph{Training procedure} The overall objective has three terms. The first term is the prediction objective. The second term is an adversarial training objective, where an auxiliary predictor $g_{\psi}$ is trained to capture label information while the feature extractor $f_{\theta_2}$ is trained to extract as little label information as possible. The third term regularizes $f_{\theta_2}$ to capture the information of a randomly selected label. For simplicity, we denote these three objective terms as $\mathcal{L}_C$, $\mathcal{L}_{A}$ and $\mathcal{L}_R$, respectively, as shown in \cref{eq:final_object}. The adversarial loss $\mathcal{L}_{A}$ and $\mathcal{L}_R$ are formulated from the goal of label protection. We reorganize the overall training objective as

{\small
\begin{equation}
\begin{aligned}
    &\theta_1, \theta_2, \theta_{\mathcal{S}}, \psi\\
    =~& \arg\min\limits_{\theta_2}\left[(1-\lambda)\min\limits_{\theta_1, \theta_S}\mathcal{L}_C+\lambda\max\limits_{\psi}\mathcal{L}_A+\lambda \mathcal{L}_R\right].
\end{aligned}
\label{eq:organized_objective}
\end{equation}
}%

We develop an algorithm of label-protecting training to optimize \cref{eq:organized_objective}, summarized in Alg.~\ref{alg:defense}. For each batch of data, we first optimize $\theta_1$ and $\theta_{\mathcal{S}}$ based on the primary task loss. Then we optimize the auxiliary predictor $\psi$. Finally, $\theta_2$ is optimized with $(1-\lambda)\mathcal{L}_C+\lambda \mathcal{L}_A + \lambda \mathcal{L}_R$. 


Note that $\theta_1, \theta_{\mathcal{S}}$ and $\psi$ are owned by the active party, and their optimization does not require additional information from the passive party except the representations $h^2$, which should be uploaded to the active party even without defense. For the passive party, the training procedure of local extractor $\theta_2$ does not change, making our defense concealed from the passive party.

\begin{algorithm}[t]  
\footnotesize
\renewcommand{\algorithmicrequire}{\textbf{Input:}}
\renewcommand{\algorithmicensure}{\textbf{Output:}}
    \caption{\textbf{Training algorithm of DIMIP.} $\textcolor{green}{\leftarrow}$ means information is sent to the active party; $\textcolor{blue}{\leftarrow}$ means information is sent to the passive party; \textcolor{red}{red steps} are conducted on the passive party.}
    \label{alg:Framwork}  
    \begin{algorithmic}[1] 
        \Require Dataset $\left\{\left(x_i^1, x_i^2, y_i\right)\right\}_{i=1}^N$; Learning rate $\eta$. 
        \Ensure $\theta_1; \theta_{\mathcal{S}}; \psi$.
        \State Initialize $\theta_1; \theta_{\mathcal{S}}; \psi$;
        \For {a batch of data $\left\{\left(x_i^1, x_i^2, y_i\right)\right\}_{i\in \sB}$}
            \State $\{H_i^2\}_{i\in\sB}\textcolor{green}{\leftarrow}\textcolor{red}{\{f_{\theta_2}\left(x_i^2\right)\}_{i\in\sB}}$;
            \State $\mathcal{L}_A\leftarrow\frac{1}{|\sB|} \sum\limits_{i\in\sB} \log g_{\psi}\left(y_i|H_i^2\right)$;
            \State $\psi \leftarrow \psi + \eta\nabla_{\psi}\mathcal{L}_A$;
            \State $\mathcal{L}_C\leftarrow\frac{1}{|\sB|} \sum\limits_{i\in\sB}\mathcal{L}\left(\mathcal{S}_{\theta_{\mathcal{S}}}\left(f_{\theta_1}\left(x_i^1\right), H_i^2\right), y_i\right)$;
            \State $\theta_1 \leftarrow \theta_1 - \eta\nabla_{\theta_1}\mathcal{L}_C$;
            \State $\theta_{\mathcal{S}} \leftarrow \theta_{\mathcal{S}} - \eta\nabla_{\theta_{\mathcal{S}}}\mathcal{L}_C$;
            \State $\{y_{k_i'}\}_{i\in\sB}\leftarrow$ randomly sample $\{y_{k_i'}\}_{i\in\sB}$ from $\{y_i\}_{i\in[N]}$.
            \State $\mathcal{L}_R\leftarrow \frac{1}{|\sB|} \sum\limits_{i\in\sB} -\log g_{\psi}\left(y_{k_i'}|H_i^2\right)$;
            \State $\{\nabla_{H_i^2}\mathcal{L}\}_{i\in\sB} \textcolor{blue}{\leftarrow} \{\nabla_{H_i^2}\left[(1-\lambda)\mathcal{L}_C+\lambda \mathcal{L}_A+\lambda \mathcal{L}_R\right]\}_{i\in\sB}$;
            \State $\textcolor{red}{{\nabla_{\theta_2}\mathcal{L} \leftarrow \frac{1}{|\sB|}\sum\limits_{i\in\sB}}\nabla_{H_i^2}\mathcal{L}\nabla_{\theta_2}{H_i^2}}$
            \State $\textcolor{red}{\theta_2 \leftarrow \theta_2 - \eta\nabla_{\theta_2}\mathcal{L}}$;
        \EndFor
        
    \end{algorithmic} \label{alg:defense}
    
\end{algorithm}

\section{Experiments}\label{sec:experiments}

\subsection{Experimental Setup}

We evaluate our proposed Party-wise Dropout and DIMIP on multiple datasets. Our evaluation focuses on two-party scenarios following the VFL literature~\cite{fu2022label, liu2019communication, liu2020federated, liu2021fate, yang2019federated}. 
\vspace{-3.5mm}
\paragraph{Baselines.} To our knowledge, Party-wise Dropout is the first approach to mitigate the performance drop after the passive party quits in VFL. But we still compare with a baseline, where the active party trains an additional head model without the quitting party to make predictions if the passive party quits, which we call \textbf{Multi-head training}. We evaluate our DIMIP against two attacks: (1) \textbf{Passive Model Completion (PMC)}~\cite{fu2022label} attack assumes that the passive party has access to an auxiliary labeled dataset. The passive party utilizes this auxiliary dataset to fine-tune a classifier that can be applied to its local feature extractor. (2) \textbf{Active Model Completion (AMC)}~\cite{fu2022label} attack is conducted by the passive party to trick the federated model to rely more on its feature extractor so as to increase its expressiveness. The passive party conducts AMC attack by actively adapting its local training configurations. We compare DIMIP with four existing defense baselines: (1) \textbf{Noisy Gradient (NG)}~\cite{chong2022baseline} is proven effective against privacy leakage in FL by adding Laplacian noise to gradients. (2) \textbf{Gradient Compression (GC)}~\cite{chong2022baseline} prunes gradients that are below a threshold magnitude, such that only a part of gradients are sent to the passive party. (3) \textbf{Privacy-preserving Deep Learning (PPDL)}~\cite{shokri2015ppdl} is a comprehensive privacy-enhancing method including three defense strategies: differential privacy, gradient compression, and random selection. (4) \textbf{DiscreteSGD (DSGD)}~\cite{chong2022baseline} conducts quantization to the gradients sent to the passive party such that the discrete gradients are used to update the adversarial party's extractor. 
\vspace{-3.5mm}
\paragraph{Datasets.} We evaluate Party-wise Dropout and DIMIP on CIFAR10 and CIFAR100. We follow \cite{liu2019communication,kang2022fedcvt,liu2021fate, yang2019federated} to split images into halves.
\vspace{-3.5mm}
\paragraph{Hyperparameter configurations.} For both CIFAR10 and CIFAR100, we use ResNet18 as backbone models. We set batch size $B$ as 32 for both datasets. We apply SGD as the optimizer with learning rate $\eta$ set to be 0.01. For DIMIP, We apply a 3-layer MLP to parameterize $g_{\psi}$. For NG defense, we apply Laplacian noise with mean of zero and scale between 0.0001-0.01.  For GC baseline, we set the compression rate from 90\% to 100\%. For PPDL, we set the Laplacian noise with scale of 0.0001-0.01, $\tau=0.001$ and $\theta$ between 0 and 0.01. For DSGD, we set the number of gradient value's levels from 1 to 2 and added Laplacian noise with the same scale as PPDL. To simulate the realistic settings in that the passive party uses different model architectures to conduct MC attacks, we apply different model architectures (MLP \& MLP\_sim) for MC attacks. The detailed model architectures can be found in Appendix~\ref{apendix:model}. The passive party has 40 and 400 labeled samples to conduct MC attacks for CIFAR10 and CIFAR100, respectively.

\vspace{-3.5mm}
\paragraph{Evaluation metrics.} (1) \textbf{Utility metric (Model accuracy):} We use the test data accuracy of the classifier on the active party to measure the performance of VFL. (2) \textbf{Robustness metric (Attack accuracy):} We use the test accuracy of the passive party's model after conducting MC attack to evaluate the effectiveness of our IP protecting method. The lower the attack accuracy, the higher the robustness against IP leakage.

\subsection{Results of Party-wise Dropout}

\begin{figure}[th]
\centering
     \includegraphics[scale=0.35]{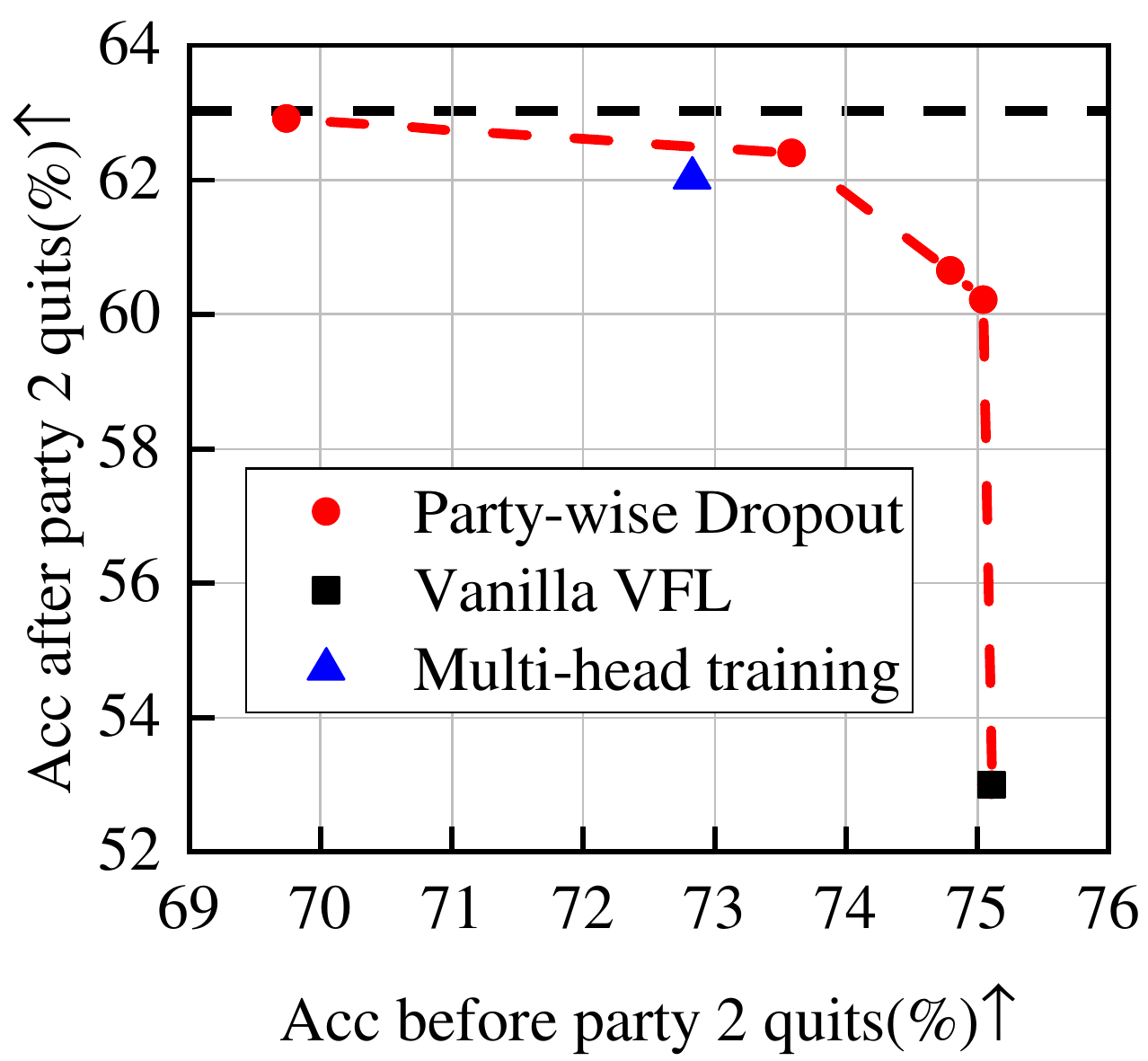}
\caption{Results of Party-wise Dropout on CIFAR10. The black dashed line denotes the accuracy of the model that party 1 trains independently.}
\label{fig:dropout_cifar10}
\end{figure}

For simplicity, we use $p$ to denote $p^2$ in \cref{eq:dropout} in our two-party experiments. We set $p$ from 0 to 0.5 to simulate the settings that the passive party has different levels of reliability. We evaluate the trade-off between the accuracy before and after the passive party quits in the deployment phase. The results of CIFAR10 are shown in \cref{fig:dropout_cifar10}, and the results of CIFAR100 are shown in \cref{tb:dropout_cifar100}. The upper bound of the test accuracy after party 2 quits is the accuracy of the model that party 1 trains independently (standalone). For CIFAR10, it is shown that applying Party-wise Dropout can improve the accuracy after party 2 quits by more than 7\% with nearly no accuracy drop before party 2 quits. By applying Party-wise Dropout, the active party can achieve nearly the same accuracy as retraining a model locally after the passive party quits by sacrificing less than 1.5\% accuracy before the passive party quits. Multi-head training can also mitigate the accuracy drop after party 2 quits. However, it cannot achieve a better trade-off than ours since our method is equivalent to multi-objective training. Notably, Multi-head training requires the active party to train $2^{K-1}$ head models according to \cref{eq:dropout}, which introduces significant computational overhead increasing rapidly with a larger $K$.

For CIFAR100, Party-wise Dropout improves the accuracy after party 2 quits by more than 5.5\% with less than 0.5\% accuracy drop before party 2 quits. Multi-head training mitigates the accuracy drop after party 2 quits, but it does not achieve a better trade-off than our method. It is shown that applying Party-wise Dropout by just setting a relatively small $p$ value can significantly improve the robustness of VFL against unexpected quitting, which shows the effectiveness of Party-wise Dropout in solving the problem of party-wise neuron co-adaptation. 

A na\"ive solution for mitigating the accuracy drop is to fine-tune the head model after a passive party quits. However, this process is time-consuming, and the service provider cannot afford to shut down the service while fine-tuning. Therefore, achieving a decent accuracy before fine-tuning is crucial. In addition, the Party-wise Dropout complements fine-tuning and can reduce computation and communication cost by starting fine-tuning from a better point.


\begin{table}[th]
\small
    \centering
    \caption{Results of Party-wise Dropout on CIFAR100.}
        \begin{tabular}{l | c | c }
            \toprule
            & \makecell[c]{Accuracy before \\party 2 quits(\%)} & \makecell[c]{Accuracy after \\party 2 quits(\%)}\\
            \hline
            $p=0$ & 44.95 & 26.65\\
            \hline
            $p=0$.05 & 44.58 & 32.01\\
            \hline
            $p=0$.1 & 44.29 & 32.03\\
            \hline
            $p=0$.3 & 42.11 & 33.05\\
            \hline
            $p=0$.5 & 40.29 & 33.85\\
            \hline
            \makecell[l]{Standalone} & N/A & 34.02\\
            \hline
            Multi-head training & 39.17 &32.72\\
            \bottomrule
        \end{tabular}
    \label{tb:dropout_cifar100}
\end{table}

\subsection{Results of DIMIP}

\begin{figure}[th]
\centering
     \includegraphics[scale=0.88]{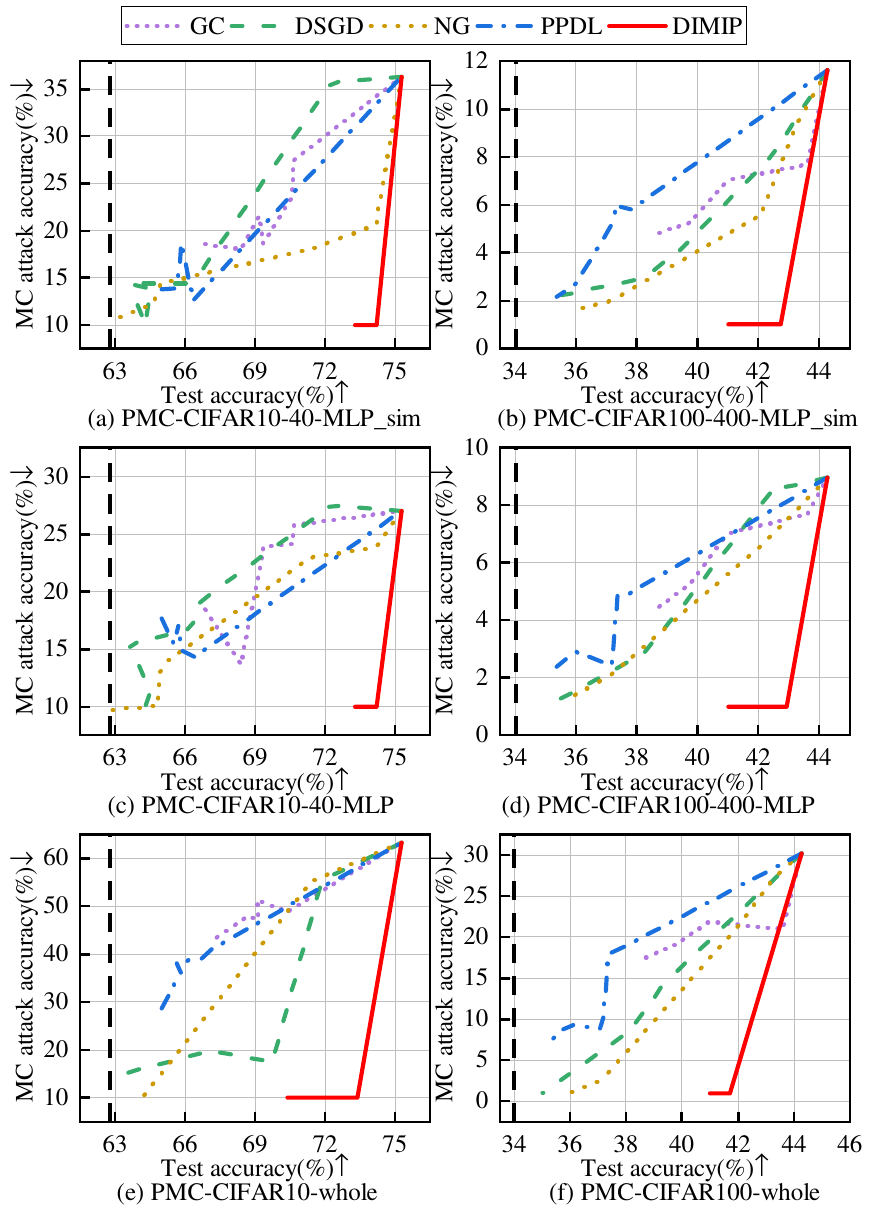}
\caption{Results of model accuracy v.s. attack accuracy on CIFAR10 and CIFAR100 against PMC attack. The black dashed line denotes the accuracy of the model that party 1 trains independently.}
\label{fig:PMC}
\end{figure}

\begin{figure}[th]
\centering
     \includegraphics[scale=0.88]{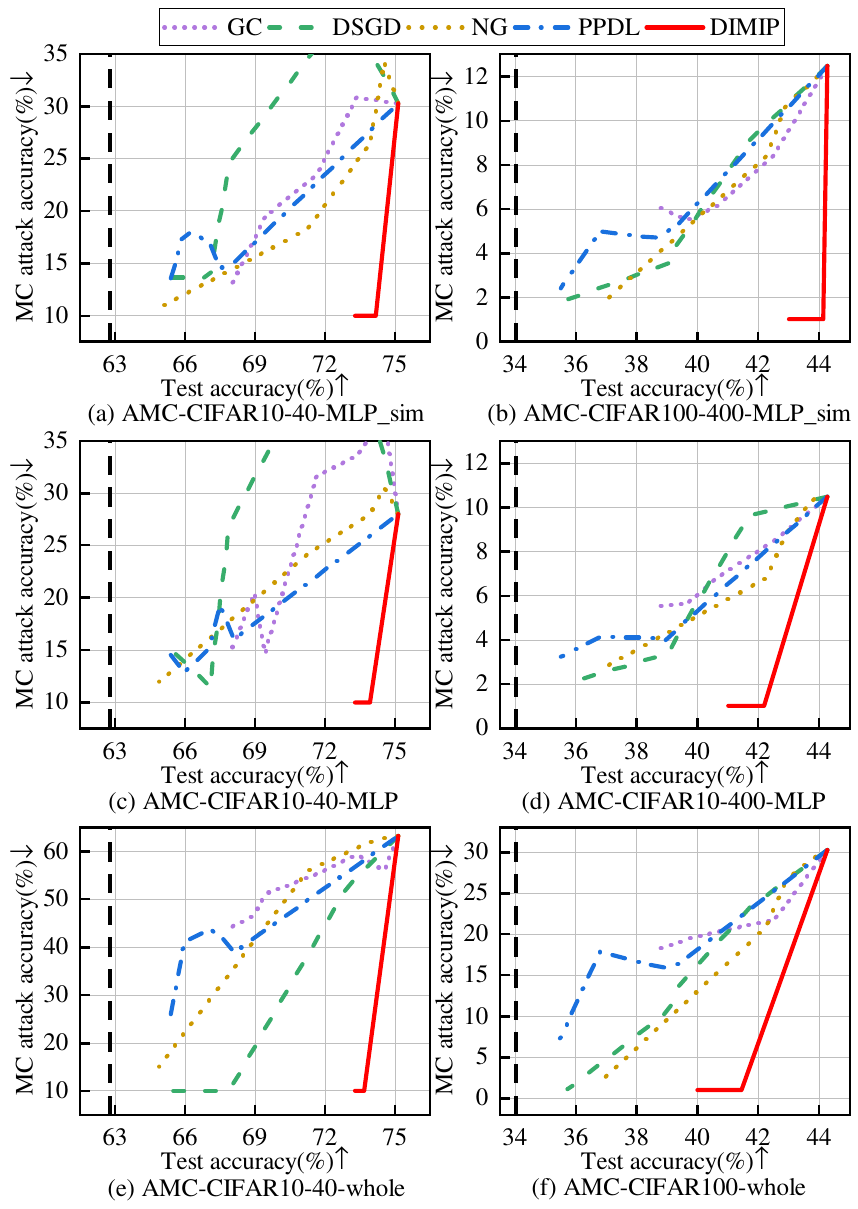}
\caption{Results of model accuracy v.s. attack accuracy on CIFAR10 and CIFAR100 against AMC attack. The black dashed line denotes the accuracy of the model that party 1 trains independently.}
\label{fig:AMC}
\end{figure}

We evaluate DIMIP on two datasets against two attack methods. We set different defense levels for our methods (i.e., different $\lambda$ values in \cref{eq:final_object}) and baselines to conduct multiple experiments to show the trade-off between the model accuracy and attack accuracy. The defense results against PMC and AMC attacks are shown in \cref{fig:PMC} and \cref{fig:AMC}, respectively. To evaluate the effectiveness of our defense in extreme cases, we also conduct experiments that the passive party has the whole labeled dataset to perform MC attacks, of which the results are shown in sub-figures (e) and (f) of \cref{fig:PMC} and \cref{fig:AMC}.

For defense against PMC on CIFAR10, our DIMIP can achieve 10\% attack accuracy (equal to random guess) by sacrificing less than 2\% model accuracy, while the other defenses drop model accuracy by more than 12\% to achieve the same defense performance. Similarly, our DIMIP can achieve 1\% attack accuracy on CIFAR100 while maintaining a model accuracy drop of less than 3\%. In contrast, the other defenses drop model accuracy by more than 9\% to achieve the same attack accuracy. Even if the passive party conducts attacks using the whole labeled training dataset, DIMIP can achieve defense performance of an attack accuracy rate of random guess with less than 3\% model accuracy drop.

Our method achieves similar results against AMC. DIMIP can achieve high defense performance of an attack accuracy rate of random guess with nearly no model accuracy drop. Notably, the other baselines improve the attack accuracy of AMC in some cases (\cref{fig:AMC}.(a) and (c)). The reason is that, by applying AMC, the model updating of the passive party is adaptive to the defense methods, making the global classifier rely more on the passive party's feature extractor.

Notably, the baselines achieve low attack accuracy only when the test accuracy degrades to nearly independent training level. This means that the baselines can only achieve strong defense performance by severely limiting the expressiveness of the passive party's feature extractor. Our method can achieve a better trade-off between the model utility and the defense performance because DIMIP only reduces the information of the true labels in the representations extracted by the passive party's feature extractor, while the general information of the data is preserved in the passive party's representations.

\subsection{Objective Analysis of DIMIP}\label{sec:L_R}

\begin{figure}[th]
\centering
     \includegraphics[scale=0.4]{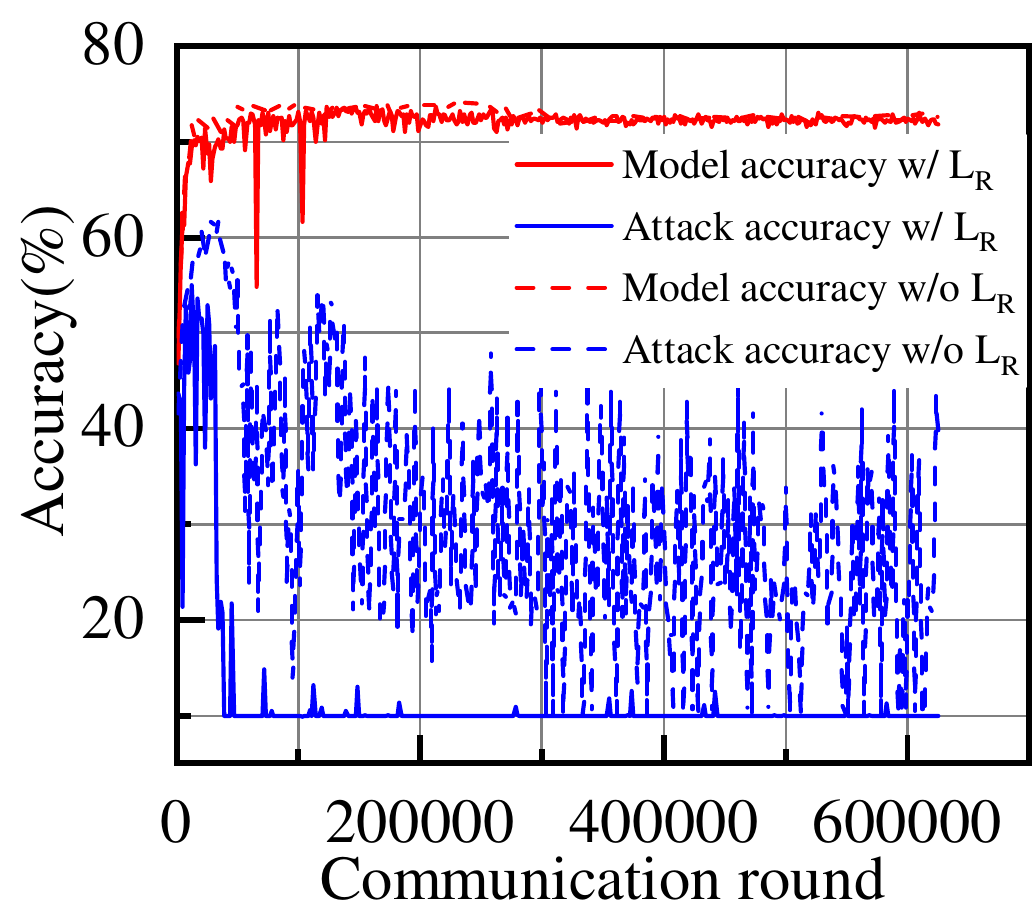}
\caption{Defense results on model accuracy and attack accuracy with and without $\mathcal{L}_R$ on CIFAR10 against PMC attack.}
\label{fig:L_R-analysis}
\end{figure}

The training objective \cref{eq:organized_objective} of DIMIP consists of 3 terms: $\mathcal{L}_C$, $\mathcal{L}_A$ and $\mathcal{L}_R$. $\mathcal{L}_C$ maintains the model utility. $\mathcal{L}_A$ is the adversarial objective to reduce the information of labels in the passive party's representations. $\mathcal{L}_R$ is also derived from the goal of mutual information reduction, but it is non-trivial to describe its functionality. To analyze the effect of $\mathcal{L}_R$, we conduct experiments that train with and without the objective $\mathcal{L}_R$ under the setting PMC-CIFAR10-whole. The results are shown in \cref{fig:L_R-analysis}. Notably, $\mathcal{L}_R$ does not influence the model accuracy, but the defense performances differ. It is shown that without $\mathcal{L}_R$, the attack accuracy can also degrade to 10\% in some communication rounds, but the degradation is much slower than training with $\mathcal{L}_R$. In addition, applying $\mathcal{L}_R$ can stabilize the defense's performance. Thus, $\mathcal{L}_R$ can boost and stabilize the performance of DIMIP defense.

\subsection{Results of Party-wise Dropout \& DIMIP}

\begin{figure}[th]
\centering
     \includegraphics[scale=0.65]{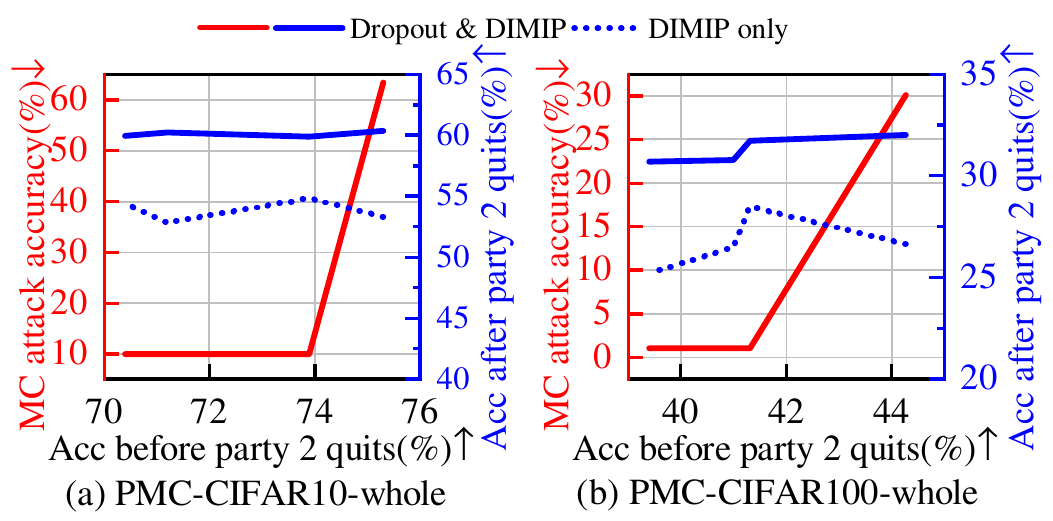}
\caption{The results of applying Party-wise Dropout and DIMIP.}
\label{fig:both}
\end{figure}

We evaluate the results of applying both Party-wise Dropout and DIMIP under PMC-CIFAR10-whole and PMC-CIFAR100-whole settings. The passive party quits in the deployment phase and tries to conduct a model completion attack using the labeled dataset. We set $p=0.05$ for Party-wise Dropout and $\lambda$ from 0 to 1 for DIMIP. The results in \cref{fig:both} show that by applying both Party-wise Dropout and DIMIP, the active party achieves 5\% higher accuracy than applying DIMIP alone if the passive party quits. Further, we prevent the passive party from achieving an attack accuracy higher than random guess levels using its feature extractor by sacrificing less than 3\% model accuracy for both datasets. Thus, our proposed Party-wise Dropout and DIMIP can improve the robustness of VFL against unexpected quitting and protect the active party's label IP effectively. 

\section{Conclusion}

We propose Party-wise Dropout to mitigate the substantial performance drop after the unexpected quitting of the passive party in VFL. We also propose a defense DIMIP against the IP leakage of the active party’s labels from the mutual information perspective with negligible degradation of the model’s utility. The experimental results show that our proposed methods can improve the robustness of VFL against unexpected quitting and protect the active party's IP effectively. In this paper, we evaluate the two-party scenario, but our theory and algorithm are naturally extendable to settings with more parties.

{\small
\bibliographystyle{ieee_fullname}
\bibliography{egbib}
}

\clearpage
\appendix
\section{Model Architecture}\label{apendix:model}

The passive party uses models with different architectures (MLP and MLP\_sim) to conduct MC attacks. MLP\_sim has one FC layer. MLP has three FC layers with a hidden layer of size $512\times256$.

\end{document}